\documentclass{jfm}

\usepackage{graphicx}
\usepackage{epstopdf,epsfig}
\usepackage{newtxtext}
\usepackage{newtxmath}
\usepackage{natbib}
\usepackage{colortbl}
\usepackage{diagbox}
\usepackage{multirow}
\usepackage{enumitem}
\usepackage{bm}

\usepackage[justification=justified]{caption}
\usepackage{hyperref}
\hypersetup{
    colorlinks = true,
    urlcolor   = blue,
    citecolor  = blue,
    linkcolor  = blue,
}

\newcommand{\RomanNumeralCaps}[1]
\nolinenumbers

\shorttitle{manuscript}
\shortauthor{Wang \& Kang}

\title{A Tug-of-War Between Baroclinic Eddies and Convection: Implications for Icy Moon Oceans}

\author{Shuang Wang\aff{1,\,2,\,3}
  \corresp{\email{shuangw@mit.edu}}
 \and Wanying Kang\aff{1}\corresp{\email{wanying@mit.edu}} \and Cheng Li\aff{2,\,3}}

\affiliation{\aff{1}Earth, Atmospheric and Planetary Science Department, Massachusetts Institute of Technology, Cambridge, MA 02139, USA \aff{2}Center for Habitable Planetary Systems, University of Michigan, Ann Arbor, MI 48109, USA \aff{3}Department of Climate and Space Sciences and Engineering, University of Michigan, Ann Arbor, MI 48109, USA}

\begin{document}

\maketitle

\begin{abstract}
In many geophysical and planetary environments, such as Earth's ocean and atmosphere as well as subsurface oceans of icy satellites, convection driven by bottom geothermal heating usually coexists with baroclinic eddies driven by lateral buoyancy/temperature gradients. These processes compete against each other, with convection destabilizing the stratification and baroclinic eddies re-stabilizing it, thereby controlling whether the bottom heat flux is significantly redistributed as it is transmitted to the upper surface. Using scaling analysis and numerical simulations, we show that a stratified layer persists near the upper surface up to ${\rm Ra}_{v}\sim {\rm Ra}_h^{5/2}$, where ${\rm Ra}_h\equiv \Delta b_0/(L_zf^2)$ measures the imposed upper-surface buoyancy contrast $\Delta b_0$ and ${\rm Ra}_v\equiv B_0/(L_z^2f^3)$ measures the strength of the bottom buoyancy flux $B_0$, $L_z$ is the domain depth and $f$ is the Coriolis parameter. 
For ${\rm Ra}_v<{\rm Ra}_h^{5/2}$, baroclinic eddies dominate over convection, maintain the upper stratified layer, and completely deflect the bottom buoyancy/heat input into meridional transport. In contrast, when ${\rm Ra}_v>{\rm Ra}_h^{5/2}$, convective plumes penetrate the stratification and transport buoyancy/heat vertically with negligible deflection. Building on these results, we further propose a scaling law for the meridional buoyancy/heat transport in this system. Applications to icy satellites are discussed.
\end{abstract}

\begin{keywords}
rotating convection, baroclinic eddy
\end{keywords}

\section{Introduction}
\label{sec:intro}
In many geophysical environments, bottom heating powered by geothermal activity commonly coexists with horizontal surface temperature gradients. In the modern Earth’s oceans and atmosphere, such gradients are sustained by differential solar radiation, whereas in subglacial lakes \citep{Wells_and_Wettlaufer_2008_Lake_lab_exps,Couston_and_Siegert_2021_LakeDynamics}  and in subsurface oceans during Snowball Earth episodes \citep{Pierrehumber_et_al_review_of_snowball} and on icy moons \citep{Kang_et_al_2022_salinity_heattransport,Kang_2023_HC_and_RBC_heatdeflection,Ames_et_al_stratification_convection} they arise from spatial variations in overlying ice thickness that modify the local freezing point. In this setting, bottom heating destabilizes the fluid interior and drives convective plumes, whereas horizontal temperature gradients induce baroclinic eddies or slantwise circulation that transport denser fluid beneath lighter fluid, thereby stratifying the upper layer \citep{Zhu-Manucharyan-Thompson-et-al-2017:influence,Lobo_et_al_2021_stable_layer,Kang_2022_different_size_heattransport,Zhang_et_al_2024_oceanweather}. Convection and baroclinic eddies thus compete to determine whether the fluid becomes stratified and how efficiently bottom heat is redistributed laterally during its upward transport through the system.

This competition has been studied in both rotating and non-rotating systems. \cite{Couston_Nandaha_Favier_2022_HC} considered a non-rotating, two-dimensional box forced by a prescribed horizontal temperature gradient at the top and a heat flux at the bottom. Their numerical simulations show that, as the vertical Rayleigh number increases, the system transitions from a strongly stratified state, characterized by a single overturning circulation confined near the upper boundary, to a convective state dominated by Rayleigh--B\'{e}nard overturning cells that penetrate the full depth. Building on this work, \cite{Rein_et_al_2025_DFD} derived an analytical scaling that captures the location of this transition.

In the presence of rotation, the single-cell overturning circulation near the upper boundary can no longer be maintained; instead, laboratory experiments show that it is replaced by small-scale eddies \citep{Wells_and_Wettlaufer_2008_Lake_lab_exps}. \cite{Callies_and_Ferrari_2018_BCIunderRBC} revisited this problem using numerical simulations and found that convection driven by bottom heating and baroclinic eddies driven by a meridional temperature gradient largely coexist. As the surface temperature gradient strengthens, baroclinic eddies intensify, enhancing restratification and suppressing convection. \cite{Kang_2023_HC_and_RBC_heatdeflection} examined this problem in the context of icy-moon oceans and proposed an analytical scaling law that predicts the minimum bottom heat flux required to overcome the stratification associated with a given surface temperature gradient. When the bottom heat flux exceeds this threshold, convective plumes can directly reach the ice shell and project the bottom heating pattern onto the ice base. Otherwise, plumes are halted by the stratification, and the accumulated heat is transported by baroclinic eddies along tilted isopycnals toward the colder surface, resulting in a substantial deflection of the bottom heating. 

A main limitation of \cite{Kang_2023_HC_and_RBC_heatdeflection} is that the study only conducted a very limited number of simulations to explore a narrow region of parameter space tailored to icy-moon conditions; it is computationally prohibitive to fully resolve convective plumes. Although the proposed heat-deflection mechanism primarily involves baroclinic eddies, it remains unclear whether adequately resolving convective plumes is essential to the problem. In particular, under-resolved convection may underestimate both convective heat transport and the ability of plumes to penetrate the stratified layer.

To fill this gap, we re-investigate the competition between convection and baroclinic eddies in a rotating, three-dimensional configuration. We explore a broad range of parameter space in which both baroclinic eddies and convective plumes are adequately resolved, and we assess the scaling proposed by \cite{Kang_2023_HC_and_RBC_heatdeflection}. \S~\ref{sec:method} describes the numerical method, governing equations, and key nondimensional parameters of the system. \S~\ref{sec:results} presents the numerical solutions and examines how the flow structure and heat-deflection behavior vary with the strength of bottom heating relative to the imposed surface temperature gradient. \S~\ref{sec:summary} provides a summary and discussion.

\section{Methods}\label{sec:method}
\begin{figure}
    \centering
    \includegraphics[width=0.7\linewidth]{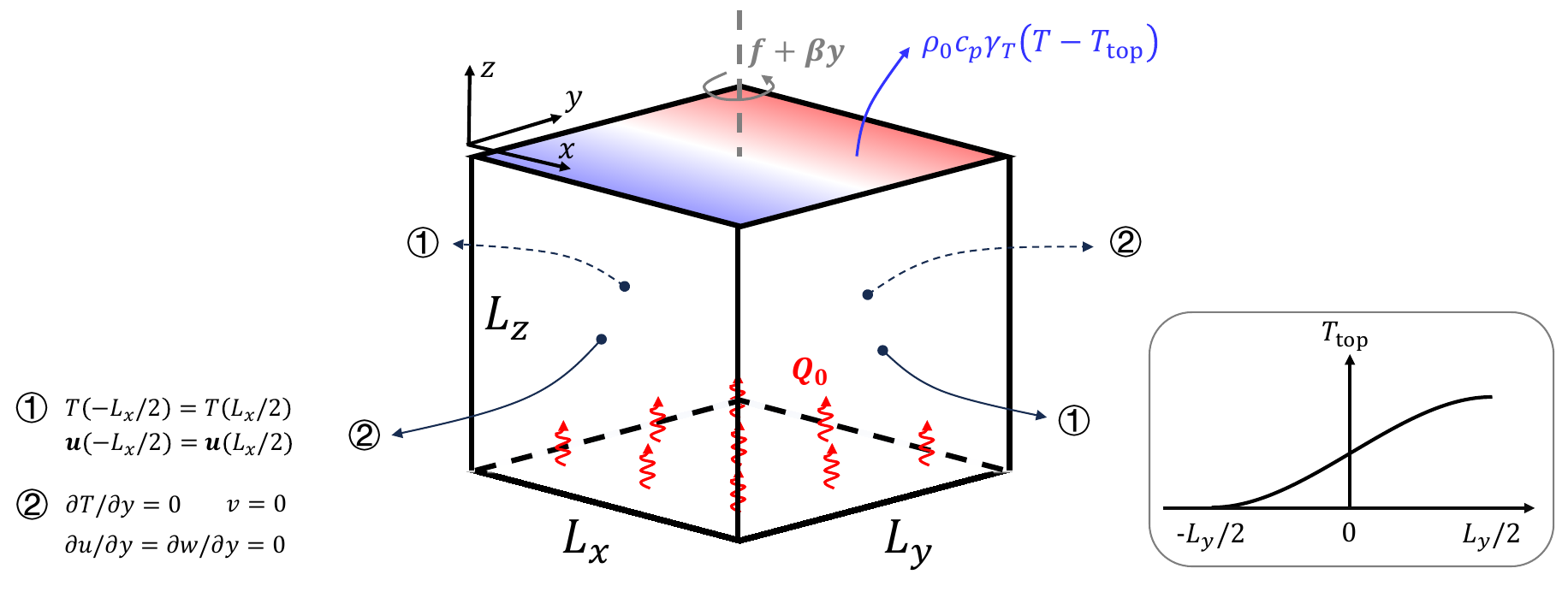}
    \caption{Sketch of the model setup. The domain has zonal, meridional, and vertical extents of $L_x$, $L_y$, and $L_z$, respectively, and rotates along $z-$axis, yielding the Coriolis parameter of $f+\beta y$. A uniform heat flux $Q_0$ (red spirals) is imposed at the bottom plate; at the top, the fluid is relaxed (blue arrow) toward an imposed sinusoidal temperature profile shown in the right panel. The lateral boundary conditions are displayed on the left.}
    \label{fig:model_diagram}
\end{figure}

To investigate the interaction between the baroclinic eddies and rotating Rayleigh--B\'{e}nard convection, we perform a suite of three-dimensional simulations in a rotating rectangular domain with zonal extent $L_x$, meridional extent $L_y$, and vertical extent $L_z$. We adopt a $\beta$-plane setup, where the Coriolis parameter $f$ varies linearly in the meridional ($y$) direction with gradient $\beta$. At the top plane, we impose a sinusoidal temperature distribution,
\begin{equation}
    T_{\rm top}=\frac{\Delta T_0}{2}\left[1+\sin\left(\frac{\pi y}{L_y}\right)\right],
\end{equation}
where $\Delta T_0$ is the imposed horizontal temperature difference. This temperature gradient is maintained by relaxing the uppermost temperature toward $T_{\rm top}$ at a sufficiently high rate $\gamma_T$. The heat exchange with the upper surface can in turn be diagnosed using $\gamma_T(T-T_{\rm top})$. At the bottom plane, a uniform heat flux $Q_0$ is prescribed to drive vertical convection. For momentum, we adopt a linear drag at the top and bottom boundaries with drag coefficient $\gamma_M$, which prevents the flow speed from growing unboundedly. The domain is periodic in the zonal direction and bounded by sidewalls in the meridional direction. No-flux and free-slip boundary conditions are set at the sidewalls (figure~\ref{fig:model_diagram}).

We use a GPU-accelerated framework \emph{Oceananigans.jl} \citep{Oceananigans_2020} to solve the non-hydrostatic Boussinesq equations, 
\begin{align}
    \boldsymbol{u}^*_{t^*}+\boldsymbol{u}^*\boldsymbol{\cdot}\boldsymbol{\nabla}^* \boldsymbol{u}^*+\boldsymbol{k}\times\boldsymbol{u}^*&=T^*\boldsymbol{k}-\boldsymbol{\nabla}^*p^*+{\rm Ek}\nabla^{*2}\boldsymbol{u}^*,\label{eq:Beq_momentum} \\
    T^*_{t^*}+\boldsymbol{u}^*\boldsymbol{\cdot}\boldsymbol{\nabla}^*T^*&={\rm Pr}^{-1}{\rm Ek}\nabla^{*2}T^*, \label{eq:Beq_temperature} \\
    \boldsymbol{\nabla}^*\boldsymbol{\cdot} \boldsymbol{u}^*&=0, \label{eq:Beq_mass}
\end{align}
where ${\rm Pr}\equiv\nu/\kappa$ is the Prandtl number, $\nu$ and $\kappa$ denote viscosity and thermal diffusivity, respectively, and ${\rm Ek}\equiv\nu/(fL_z^2)$ is the Ekman number. Here and hereafter, superscript stars denote nondimensional quantities. Detailed nondimensionalization is provided in the Supplementary Material (SM). 

In this system, six nondimensional numbers control the dynamics, including $\rm Pr$, $\rm Ek$, two aspect ratios $\Gamma_x\equiv L_x/L_z$ and $\Gamma_y\equiv L_y/L_z$, and the horizontal and vertical Rayleigh numbers, 
\begin{equation}
    {\rm Ra}_h\equiv\frac{g\alpha\Delta T_0}{f^2L_z},\quad {\rm Ra}_v\equiv\frac{g\alpha Q_0}{\rho_0c_pf^3L_z^2},
\end{equation}
where $g$ is the gravity, $\alpha$ is the thermal expansivity, $\rho_0$ is the reference density of the fluid, and $c_p$ is the specific heat capacity.
The two Rayleigh numbers are involved in the top and bottom boundary conditions:
\begin{align}
    Q^*_{v,{\rm top}}&=\gamma_T^*\left\{T^*-\frac{{\rm Ra}_h}{2}\left[1+\sin\left(\pi y^*\Gamma_y^{-1}\right)\right]\right\} &(z^*=1), \\
    Q^*_{v,{\rm bot}}&={\rm Ra}_{v} &(z^*=0).
\end{align}

The definitions of ${\rm Ra}_h$ and ${\rm Ra}_v$ are independent of $\nu$ and $\kappa$. The parameter ${\rm Ra}_v$ is usually referred to as the modified flux Rayleigh number \citep{Christensen_2002_rotating_convection,Aurnou_et_al_2020_scaling_for_convection}. These Rayleigh numbers can be related to the conventional (flux) Rayleigh numbers through 
\begin{equation}
    {\rm Ra}_h=\frac{{\rm Ra}_h^{\dag}{\rm Ek}^2}{\rm Pr},\quad {\rm Ra}_v=\frac{{\rm Ra}_F{\rm Ek}^3}{{\rm Pr}^2}=\frac{{\rm Ra}_v^{\dag}{\rm Nu}\,{\rm Ek}^3}{{\rm Pr}^2},
\end{equation}
where ${\rm Ra}_F\equiv(g\alpha Q_0L_z^4)/(\rho_0c_p\nu\kappa^2)$ is the flux Rayleigh number, ${\rm Ra}_h^{\dag}\equiv(g\alpha \Delta T_0L_z^3)/(\nu\kappa)$ and ${\rm Ra}_v^{\dag}\equiv(g\alpha \Delta_v TL_z^3)/(\nu\kappa)$ are Rayleigh number based on the horizontal temperature contrast $\Delta T_0$ and the vertical temperature contrast $\Delta_vT$, respectively, and ${\rm Nu}\equiv(Q_0L_z)/(\rho_0c_p\kappa\Delta_vT)$ is the Nusselt number \citep{Rein_et_al_2025_DFD}.  

These two Rayleigh numbers can also be expressed as
\begin{equation}
    {\rm Ra}_h=\left(\frac{L_{d,{\rm max}}}{L_z}\right)^2,\quad {\rm Ra}_v=\left(\frac{L_{\rm rot}}{L_z}\right)^2.
\end{equation}
Here, $L_{d,{\rm max}}\equiv\sqrt{\Delta b_0L_z}/f$ is a Rossby deformation radius, where $\Delta b_0=g\alpha\Delta T_0$ is the buoyancy contrast at the top surface \citep{Pedlosky_1987,Vallis_2006}, and $L_{\rm rot}\equiv\sqrt{B_0/f^3}$ is the characteristic length of a rotating convective plume, where $B_0=g\alpha Q_0/(\rho_0c_p)$ is the bottom buoyancy flux \citep{Jones_and_Marshall_1993_openoceanconvection,Maxworthy_and_Narimousa_1994_largerotatingconvection}. Therefore, ${\rm Ra}_h$ describes the relative size of a baroclinic vortex to the domain extent, and ${\rm Ra}_v$ describes whether a plume is constrained by rotation when it reaches the top surface.

Our simulations do not lie in the diffusivity-free regime for rotating convection \citep[${\rm Ra}\lesssim{\rm Ek}^{-8/5}$,][]{Gastine_Wicht_Aubert_2016}, implying the presence of thin boundary layers influenced by $\nu$ and $\kappa$. Nevertheless, the convective dynamics in the interior are fully controlled by the imposed heat flux $Q_0$. Because the interaction between convection and baroclinic eddies occurs in the interior, the buoyancy jump across the bottom boundary layer has little dynamical significance. Moreover, the buoyancy jump within the upper boundary layer is suppressed by the large $\gamma_T$, which may be physically interpreted as enhanced boundary-layer turbulent mixing driven by wind stress on Earth or by tidal ice libration on icy satellites. Under this setup, the dynamics of interest are expected to be governed by diffusivity-free parameters ${\rm Ra}_h$, ${\rm Ra}_v$, $\Gamma_x$, and $\Gamma_y$. As a sanity check, we verified that all simulations follow the diffusivity-free ${\rm Nu}$--${\rm Ra}$ relationship.

Fully surveying the parameter space spanned by four nondimensional numbers remains a formidable task. We therefore focus on how heat-deflection behavior varies across the parameter space defined by ${\rm Ra}_h$ and ${\rm Ra}_v$, while fixing the aspect ratios to $\Gamma_x=3$ and $\Gamma_y=4$ by default. We also conduct a series of simulations with $\Gamma_x=\Gamma_y=1$ to examine the influences of aspect ratio on the heat deflection. The configurations of all simulations are listed in table~\ref{tab:we_parameter}, and the rationale for the chosen parameter ranges are provided in the SM.

\begin{table}
    \centering
    \renewcommand{\arraystretch}{1.1}
    \begin{tabular}{lccccc}
        \multicolumn{5}{c}{Group-1} \\ 
        ${\rm Ra}_h$ & \multicolumn{4}{c}{0.04} & \\ 
        {${\rm Ra}_v$} & \multicolumn{4}{c}{$1\times10^{-8}$, $4\times10^{-8}$, $1\times10^{-7}$, $4\times10^{-7}$, $1\times10^{-6}$, $4\times10^{-6}$, $1\times10^{-5}$, $4\times10^{-5}$} & \\
        ${\rm Ek}$ & \multicolumn{4}{c}{$1.1\times10^{-7}$} & \\
        \hline
        \multicolumn{5}{c}{Group-2} \\ 
        ${\rm Ra}_h$ & \multicolumn{5}{c}{0.02, 0.04, 0.1, 0.2} \\ 
        ${\rm Ra}_v$ & \multicolumn{4}{c}{$4\times10^{-6}$, $1\times10^{-5}$, $4\times10^{-5}$, $1\times10^{-4}$, $4\times10^{-4}$, $1\times10^{-3}$} \\ 
        ${\rm Ek}$ & \multicolumn{5}{c}{$4\times10^{-9}$} \\
    \end{tabular}
    \caption{Parameters used in all simulations. Common setup includes ${\rm Pr}=1$ and $\Delta f/f=\beta L_y/f\approx0.5$. In Group-1, $\Gamma_x=3$ and $\Gamma_y=4$, and $\gamma_T^*=0.33$. In Group-2, $\Gamma_x=\Gamma_y=1$, and $\gamma_T^*=0.02$.}
    \label{tab:we_parameter}
\end{table}

\section{Results}\label{sec:results}


\begin{figure}
    \centering
    \includegraphics[width=1\linewidth]{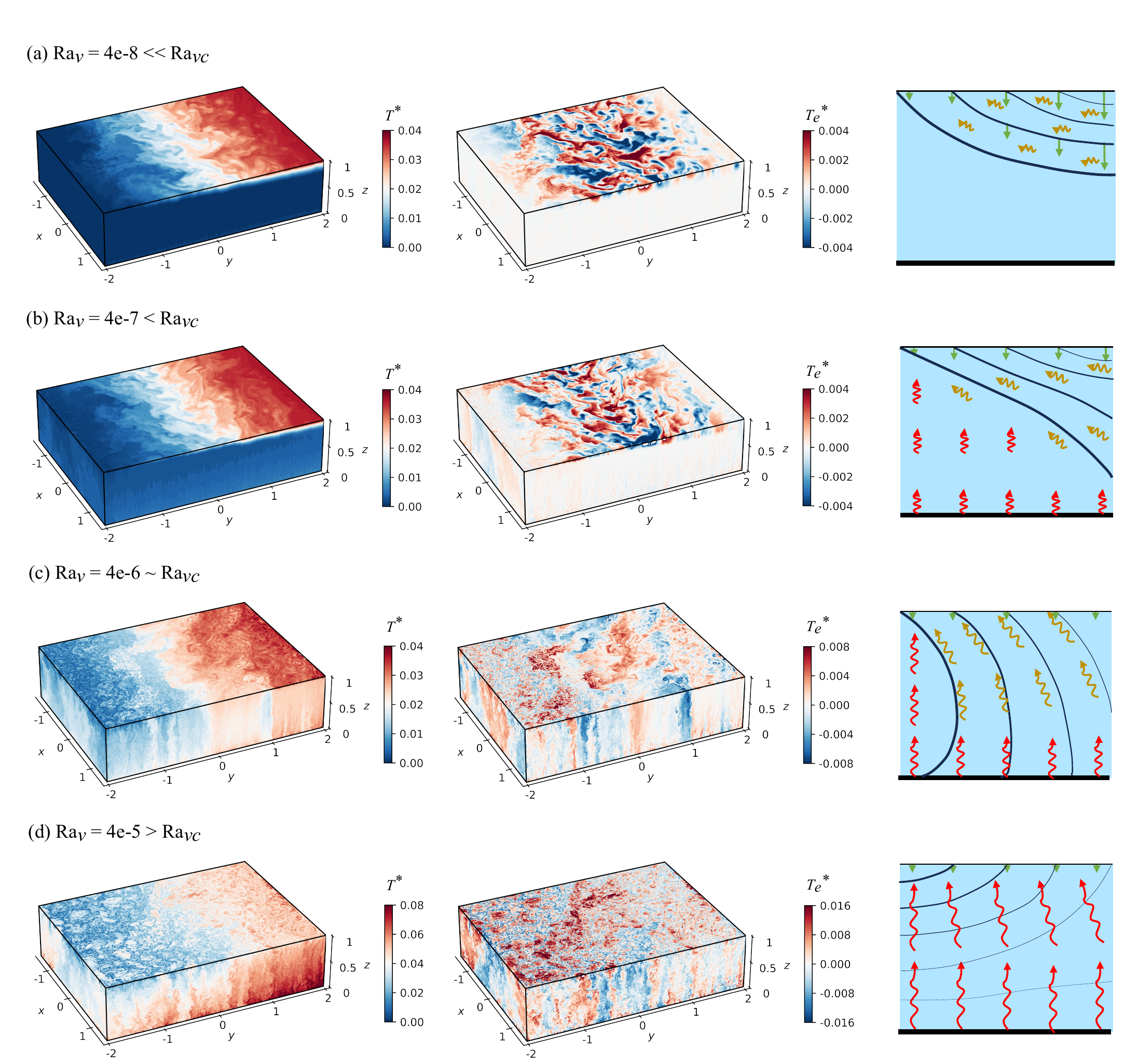}
    \caption{Three-dimensional fields of nondimensional temperature $T^*$ (left column) and corresponding anomaly $T_e^*$ after subtracting the zonal mean (middle column), in the Group-1 simulations with ${\rm Ra}_h=0.04$ and (a) ${\rm Ra}_v=4\times10^{-8}$, (b) ${\rm Ra}_v=4\times10^{-7}$, (c) ${\rm Ra}_v=4\times10^{-6}$, and (d) ${\rm Ra}_v=4\times10^{-5}$. The right column shows schematic diagrams of the shape of isopycnals (black solid curves) and heat transport components, including downward diffusion (green vectors), baroclinic transport (yellow spirals), and convective transport (red spirals).}
    \label{fig:b_field}
\end{figure}

\begin{figure}
    \centering
    \includegraphics[width=1\linewidth]{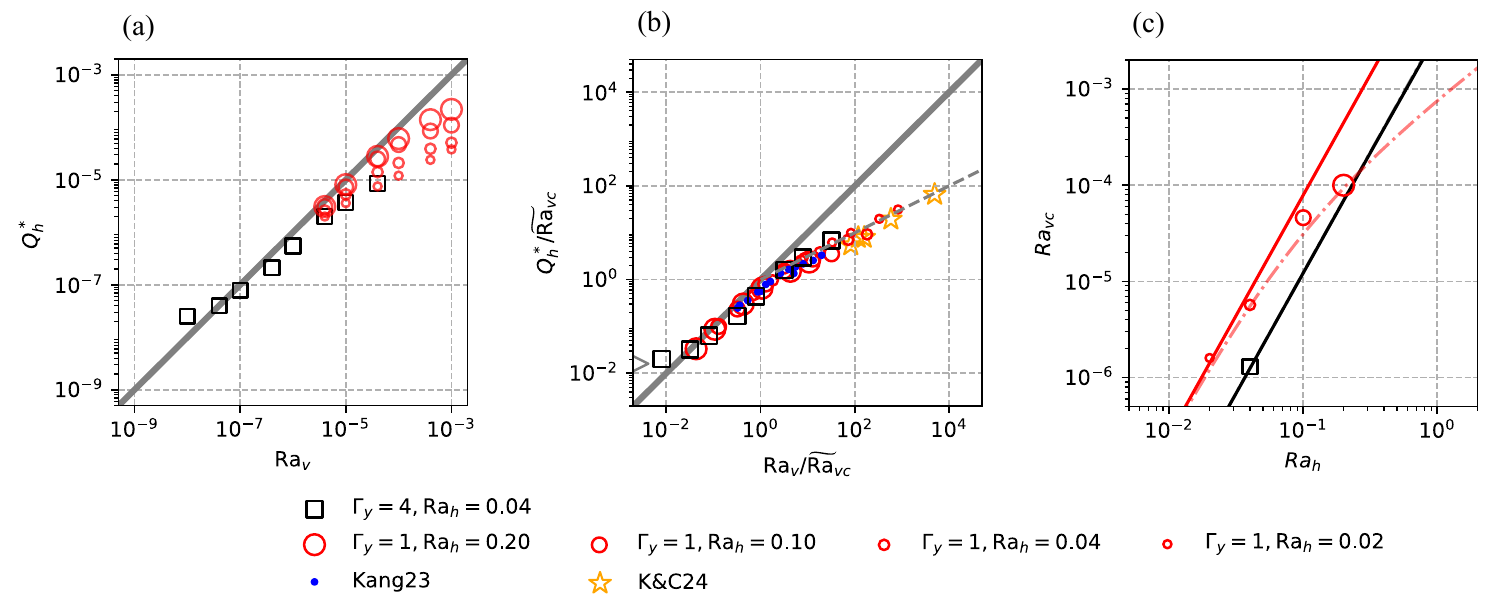}
    \caption{(a) Meridional heat flux $Q_h^*=2\overline{|\iint (vT)^*\mathrm{d}x^*\mathrm{d}z^*|}/(\Gamma_x\Gamma_y)$ against the vertical Rayleigh number ${\rm Ra}_{v}$. (b) Similar to (a), but both $Q_h^*$ and ${\rm Ra}_v$ are rescaled by $\widetilde{\rm Ra}_{vc}$. (c) ${\rm Ra}_{vc}$ against ${\rm Ra}_h$. In all panels, black squares represent simulations in Group-1; gray triangles mark $Q_h^*$ without bottom heating; red circles represent simulations in Group-2; blue dots represent results from \cite{Kang_2023_HC_and_RBC_heatdeflection}, labeled as Kang23; and yellow stars represent results from \cite{Kvorka_and_Cadek_2024_DALeffect}, labeled as K\&C24. In panels (a) and (b), gray solid and dashed lines represent the one-to-one line and the best-fit relation $y=x^{1/2}$, respectively. In panel (c), solid lines represent predicted ${\rm Ra}_{vc}$ from equation~(\ref{eq:Ravc}) and dashed curves represent $\widetilde{\rm Ra}_{vc}$ from equation~(\ref{eq:Ravc_modified}). For Group-1 (black), ${\rm Ra}_{vc}$ and $\widetilde{\rm Ra}_{vc}$ overlap.}
    \label{fig:heat_deflection}
\end{figure}


Figure \ref{fig:b_field} presents the full buoyancy field and its anomaly after subtracting the zonal mean for a series of simulations with fixed ${\rm Ra}_h$. As ${\rm Ra}_v$ increases, the solution transitions from a baroclinic-eddy-dominated state to a convection-dominated state, consistent with previous studies \citep{Callies_and_Ferrari_2018_BCIunderRBC,Kang_2023_HC_and_RBC_heatdeflection}.     

When bottom heating is negligible, the flow consists solely of baroclinic eddies. These eddies extract the available potential energy of the system by transporting buoyant fluid upward and dense fluid downward along isopycnals, thereby flattening their slope. This flattening is balanced by downward heat diffusion from the surface. In equilibrium, a stratified layer of finite depth forms beneath the warm side of the domain, where isopycnals originating from the upper surface bend downward and outcrop at the sidewall (figure~\ref{fig:b_field}a). Heat diffused downward from the warm side of the surface is returned upward along isopycnals by baroclinic eddies and absorbed at the cold side of the surface.

As ${\rm Ra}_v$ increases modestly, convection begins to emerge, although the flow remains dominated by baroclinic eddies (figure~\ref{fig:b_field}b). Over most of the domain, convection is inhibited by the stratification generated by baroclinic eddies, except within a very narrow band near the cold wall, where stratification is absent (left panel in figure~\ref{fig:b_field}b). Convective plumes are readily distinguishable from baroclinic eddies by their smaller horizontal scale and sharper temperature-anomaly structure (mid panel in figure~\ref{fig:b_field}b). This scale separation is expected because the (nondimensional) Rossby deformation radius $L_d^*=\sqrt{{\rm Ra}_h\Gamma_D}$ ($\Gamma_D$ denotes the nondimensional penetration depth of the stratified layer diagnosed from the simulations), which characterizes the baroclinic eddy scale, is at least ten times larger than the characteristic size of convective plumes when they reach the upper boundary, which scales as $L_{\rm cone}^*={\rm Ra}_v^{1/4}$ \citep{Jones_and_Marshall_1993_openoceanconvection,Maxworthy_and_Narimousa_1994_largerotatingconvection}. In this regime, bottom heating is first carried upward by convection, which transports heat across isopycnals. Upon encountering the stratified layer, baroclinic eddies take over and transport heat upward, which channels the heat along tilted isopycnals \citep{Jayne_and_Marotzke_2002_OceanicEddyHeatTransport,Vallis_2006,Kang_2023_HC_and_RBC_heatdeflection,Zhang_et_al_2024_oceanweather}. Because these isopycnals tilt upward from the warm side toward the cold side, only a small fraction of the cold-side surface receives heat originating directly from beneath (figure~\ref{fig:b_field}b).

With further increase in ${\rm Ra}_v$, more plumes penetrate the stratified layer, and the horizontal area occupied by convective plumes expands accordingly (mid panel in figure~\ref{fig:b_field}c). As the stratification is eroded by convection, some isopycnals originating from the upper surface are drawn downward and eventually reach the bottom of the domain (left panel in figure~\ref{fig:b_field}c). Along these bottom-touching isopycnals, baroclinic eddies transport heat to the upper surface, increasing the fraction of the surface that directly receives bottom heating. At sufficiently large ${\rm Ra}_v$, the system transitions into a convection-dominated regime, where convection spans the entire domain, the surface stratification observed at lower ${\rm Ra}_v$ completely disappears, and bottom heating is directly projected onto the upper surface (figure~\ref{fig:b_field}d). 

To quantify heat-deflection efficiency, we diagnose the nondimensional meridional heat flux $Q_h^*$ in each simulation and plot it as a function of ${\rm Ra}_v$ in figure~\ref{fig:heat_deflection}a. For each fixed ${\rm Ra}_h$, $Q_h^*$ remains nearly a constant at $Q_{h\kappa}^*$ when ${\rm Ra}_v$ is below a critical value ${\rm Ra}_{v\kappa}$. In this regime, convection is too weak to influence the upper stratified layer, and heat transport depends primarily on ${\rm Ra}_h$ and ${\rm Ek}$, rather than on ${\rm Ra}_v$. Beyond ${\rm Ra}_{v\kappa}$, $Q_h^*$ increases with ${\rm Ra}_v$ along a one-to-one line, indicating that the meridional heat transport equals the bottom heat input and complete deflection of bottom heating toward the cold side of the surface. As ${\rm Ra}_v$ increases beyond a threshold ${\rm Ra}_{vc}$, $Q_h^*$ begins to fall below the one-to-one line, implying that an increasing fraction of bottom heat flux is projected directly onto the upper surface. The critical Rayleigh number ${\rm Ra}_{vc}$, defined as the value at which $Q_h^*/{\rm Ra}_v$ first drops below $0.5$, is plotted as a function of ${\rm Ra}_h$ in figure~\ref{fig:heat_deflection}c. As expected, ${\rm Ra}_{vc}$ increases with ${\rm Ra}_h$, since a larger imposed surface buoyancy contrast produces stronger stratification and therefore requires a larger bottom heat flux to overcome.

To understand what sets the bottom heat flux required to overcome the surface stratification, i.e., ${\rm Ra}_{vc}$, we follow \cite{Kang_2023_HC_and_RBC_heatdeflection}, who proposed that the transition occurs when isopycnals originating from the upper surface just reach the bottom boundary. Using baroclinic turbulence scaling \citep{Held_and_Larichev_1996_BCIscaling,Jansen_and_Ferrari_2013_BCIscaling,Kang_2022_different_size_heattransport,Kang_2023_HC_and_RBC_heatdeflection,Zhang_et_al_2024_oceanweather}, one can estimate the upward eddy heat transport $Q_v^*$ for any given isopycnal slope $s\equiv\Gamma_D\Gamma_y^{-1}$
\begin{equation}\label{eq:Hv_slope}
    Q_v^*=k\Gamma_y^{-\frac{1}{2}}s^{\frac{5}{2}}{\rm Ra}_h^{\frac{5}{2}},
\end{equation}
where $k\approx0.25$ is a fitted constant \citep{Vallis_2006}. When the isopycnal slope $s=\Gamma_y^{-1}$, the critical bottom heat flux $Q_{vc}^*$ that sustains such an isopycnal slope must balance this upward eddy heat transport. In nondimensional form, this balance yields the critical vertical Rayleigh number (see SM)
\begin{equation}\label{eq:Ravc}
{\rm Ra}_{vc}=k\Gamma_y^{-3}{\rm Ra}_h^{\frac{5}{2}}.
\end{equation}
In our simulations, however, the temperature beneath the upper boundary is relaxed toward an imposed temperature profile $T_{\rm top}$ at a rate of $\gamma_T$ (units of m\,s$^{-1}$). With a finite $\gamma_T$, a temperature drop will always establish in the uppermost layer, especially when the heat flux is strong. This temperature drop increases from the warm side toward the cold side because more heat tends to be delivered toward the latter. Consequently, the horizontal temperature contrast in the fluid interior is reduced relative to what is set at the upper boundary, i.e., $\Delta T_0$, and thereby the effective ${\rm Ra}_h$ and ${\rm Ra}_{vc}$. Accounting for this effect, the modified $\widetilde{\rm Ra}_{vc}$ can be solved from
\begin{equation}\label{eq:Ravc_modified}
    \widetilde{\rm Ra}_{vc}=k\Gamma_y^{-3}\left({\rm Ra}_h-\frac{2\widetilde{\rm Ra}_{vc}}{\gamma_T^*}\right)^{\frac{5}{2}}.
\end{equation}
$\widetilde{\rm Ra}_{vc}$ is always smaller than ${\rm Ra}_{vc}$ (red solid line and dashed curve in figure~\ref{fig:heat_deflection}c) and approaches ${\rm Ra}_{vc}$ in the limit of large $\gamma_T^*$ (black solid line and dashed curve in figure~\ref{fig:heat_deflection}c). The detailed derivation is provided in the SM.

We normalize both axes in figure~\ref{fig:heat_deflection}a with $\widetilde{\rm Ra}_{vc}$ to obtain figure~\ref{fig:heat_deflection}b, such that the abscissa ${\rm Ra}_v/\widetilde{\rm Ra}_{vc}$ measures the relative dominance of convection over baroclinic eddies according to the above threshold. Under this rescaling, data obtained at different ${\rm Ra}_h$ and $\Gamma_y$ roughly collapse onto a single curve (figure~\ref{fig:heat_deflection}b). As ${\rm Ra}_v/\widetilde{\rm Ra}_{vc}$ increases beyond unity, data start to deviate from the one-to-one line and approximately follow a best-fit one-half power law.

We collect data from previous studies that consider similar configurations \citep{Kang_2023_HC_and_RBC_heatdeflection, Kvorka_and_Cadek_2024_DALeffect} and overlay their results in figure \ref{fig:heat_deflection}b to compare with ours. Details of the compilation are provided in the SM. The dataset from \cite{Kang_2023_HC_and_RBC_heatdeflection} lies close to $\widetilde{\rm Ra}_{vc}$ and overlaps well with our results. The dataset from \cite{Kvorka_and_Cadek_2024_DALeffect}, who extrapolate their results to Titan (Saturn's sixth icy moon), spans ${\rm Ra}_{v}/{\rm Ra}_{vc}$ from $50$ to $5000$ and falls along the extension of the scaling curve inferred from our results, suggesting broad consistency across parameter regimes. However, the parameter choice in \cite{Kvorka_and_Cadek_2024_DALeffect} overestimates Titan's ${\rm Ra}_v/{\rm Ra}_{vc}$ by 7--8 orders of magnitude, which casts doubt on the direct relevance of the results.


We next seek to understand the transition from diffusion-powered baroclinic eddies to convection-powered baroclinic eddies, demarcated by ${\rm Ra}_{v\kappa}$. In the diffusion-dominated regime, the equilibrium isopycnal slope $s$ is set by the balance between upward eddy heat transport and downward diffusion, namely
\begin{align}\label{eq:heatbalance_kappa}
    Q_{v\kappa}=\rho_0c_p\kappa\frac{\Delta T_0}{D_\kappa}\;{\rm (dimensional)}\;\rightarrow\;
    Q_{v\kappa}^*={\rm Pr}^{-1}{\rm Ek}\,\Gamma_{D\kappa}^{-1}{\rm Ra}_h\;(\rm nondimensional)
\end{align}
where $D_\kappa$ is the penetration depth of the stratified layer without bottom heating, $\Gamma_{D\kappa}$ denotes its nondimensional form, and the subscript `$\kappa$' refers to the diffusion-powered regime. Combining equations~(\ref{eq:Hv_slope}) and (\ref{eq:heatbalance_kappa}), we can solve $\Gamma_{D\kappa}$ and  $Q_{v\kappa}^*$ (equal to the horizontal transport $Q_{h\kappa}^*$; see SM for details):
\begin{align}\label{eq:D_and_Qkappa}
    \Gamma_{D\kappa}=k^{-\frac{2}{7}}{\rm Pr}^{-\frac{2}{7}}{\rm Ek}^{\frac{2}{7}}\Gamma_y^{\frac{6}{7}}{\rm Ra}_h^{-\frac{3}{7}},\quad 
    Q_{v\kappa}^*=k^{\frac{2}{7}}{\rm Pr}^{-\frac{5}{7}}{\rm Ek}^{\frac{5}{7}}\Gamma_y^{-\frac{6}{7}}{\rm Ra}_h^{\frac{10}{7}}.
\end{align}
These scalings agree with previous results \citep{Kang_2022_different_size_heattransport,Kang_2023_HC_and_RBC_heatdeflection,Zhang_et_al_2024_oceanweather}. The value of ${\rm Ra}_{v\kappa}$ is therefore set by $Q_{v\kappa}^*$. For ${\rm Ra}_v<Q_{v\kappa}^*$, the horizontal heat flux associated with slantwise convection is negligible compared with the diffusive contribution. For ${\rm Ra}_v>Q_{v\kappa}^*$, $Q_h^*$ becomes predominantly convection-induced, as evidenced by its one-to-one scaling with ${\rm Ra}_v$. For each fixed ${\rm Ra}_{h}$, we present the predicted $Q_{h\kappa}^*$ in figure~\ref{fig:heat_deflection} using gray triangles, and it well matches the heat transport in the small-${\rm Ra}_{v}$ limit.

\section{Conclusion and discussion}\label{sec:summary}
In this study, we investigated fluid dynamics in a system simultaneously forced by bottom heating and a surface temperature gradient, with a focus on the competition between baroclinic eddies and Rayleigh--B\'{e}nard convection. Forced by horizontal temperature gradients, baroclinic eddies subduct dense fluid beneath buoyant fluid, creating a near-surface stratified layer that suppresses convection (figure~\ref{fig:b_field}a). As bottom heating increases, the system undergoes a sequence of regimes: bottom heat is initially completely deflected toward the colder side of the surface by the stratification (figure~\ref{fig:b_field}b); convection then progressively erodes the stratified layer (figure~\ref{fig:b_field}c); and eventually the stratification disappears, allowing direct heat transport from the bottom to the surface (figure~\ref{fig:b_field}d).

A key question is what controls the onset of this transition. We tested the analytical scaling law for the critical bottom heat flux proposed by \cite{Kang_2023_HC_and_RBC_heatdeflection} (see also equation~\ref{eq:Ravc}) using numerical simulations spanning a broad parameter range. Unlike previous studies, our simulations are not tailored to a specific planetary body but instead to ensure that both convection and baroclinic eddies are adequately resolved. The proposed scaling law agrees well with the numerical results (figure~\ref{fig:heat_deflection}). In addition, we find that as bottom heating approaches zero, the horizontal heat transport remains finite and is set by diffusion. Analytical scaling laws for both the transition threshold and heat transport (equations~\ref{eq:Ravc}, \ref{eq:Ravc_modified}, and \ref{eq:D_and_Qkappa}) are derived and validated. Our work thus provides a predictive framework for determining whether a rotating fluid is stratified or convective and how heat may be redistributed meridionally. 

\begin{table}
    \centering
    \renewcommand{\arraystretch}{1.1}
    \setlength{\tabcolsep}{4.5pt}
    \begin{tabular}{lccc}
        & Europa & Enceladus & Titan \\
        \hline
        $\Delta D_{\rm ice}$ (km) & $0.1\sim2$ & $15\sim25$ & $1\sim10$ \\
        ${\rm Ra}_h$ & $0.05\sim1.0$ & $0.01\sim0.02$ & $4.6\sim46.4$ \\
        ${\rm Ra}_v$ & $\leq2.3\times10^{-8}$ & $\leq4.4\times10^{-10}$ & $\leq3.7\times10^{-7}$ \\
        ${\rm Ra}_{vc}$ & $5.6\times10^{-8}\sim1.0\times10^{-4}$ & $4.4\times10^{-9}\sim1.6\times10^{-8}$ & $0.01\sim4.47$ \\
        ${\rm Ra}_v/{\rm Ra}_{vc}$ & $\leq0.4$ & $\leq0.1$ & $\leq4\times10^{-5}$
    \end{tabular}
    \caption{Estimated parameter values for Europa, Enceladus, and Titan. Other parameters can be found in table~5 in the SM. For all three satellites, we assume the maximum global mean bottom heat flux $Q_0$ to be about 40~mW\,m$^{-2}$.}
    \label{tab:icy_moon_regime}
\end{table}

These findings have important implications for icy-moon oceans. Thus far, three icy moons, including Europa (Jupiter's second icy moon), Enceladus  (Saturn's second icy moon), and Titan (Saturn's sixth icy moon), have some constraints on their ice-thickness variations, which allow us to estimate their ${\rm Ra}_h$ and ${\rm Ra}_{vc}$. The bottom heating rates on icy moons are not directly measured, but they are constrained by the maximum heat that can be conducted through the ice shell and radiated to space, which should not exceed $40$~mW~m$^{-2}$ for all three cases. Based on these estimates (see table \ref{tab:icy_moon_regime}), we arrive at the conclusion that all three icy moons should have a stratified upper ocean that deflects all bottom heating toward the latitudes with thicker ice shell, consistent with \cite{Kang_2023_HC_and_RBC_heatdeflection}. Since bottom heating cannot reach the thinner part of ice shell, the observed large-scale ice-thickness variations on these icy moons are unlikely to be sustained by heating generated in the silicate core alone, but instead require spatially heterogeneous heating within the ice shell. This framework may also be applied to atmospheres of gas giants to help explain the latitudinal distribution of outgoing longwave radiation from the planet, given that these atmospheres are forced by internal heating from below and heterogeneous solar heating from above.

\backsection[Acknowledge]{We thank MIT Engaging, NSF ACCESS, NCAR Derecho, and NCSA Delta for support with computing resources, and MIT Svante for support with data storage.}

\backsection[Funding]{WK acknowledges support from MIT startup funding and NASA ICAR award 80NSSC26K0263. CL acknowledges support from NASA Juno project NNM06AA75C and subaward to the University of Michigan with project No. Q99063JAR.}

\backsection[Declaration of interests] {The authors report no conflict of interest.}

\backsection[Data availability statement]{All the data supporting this work are available from the corresponding author upon reasonable request.}

\backsection[Author ORCIDs]{Shuang Wang, https://orcid.org/0000-0002-9406-1781; Wanying Kang, https://orcid.org/0000-0002-4615-3702; Cheng Li, https://orcid.org/0000-0002-8280-3119}




\bibliographystyle{jfm}
\bibliography{jfm}


\end{document}